\documentclass[aps,pre,reprint,groupedaddress]{revtex4-1}

\usepackage[dvipdfmx]{graphicx}
\usepackage{amsmath, amssymb}
\usepackage[normalem]{ulem}
\usepackage{color}
\usepackage{tabularx}

\begin{document}


\title{Lifetime distributions for adjacency relationships in a Vicsek model}


\author{Takuma Narizuka}
\email[]{pararel@gmail.com}
\affiliation{Department of Physics, Faculty of Science and Engineering, Chuo University, Bunkyo, Tokyo 112-8551, Japan}

\author{Yoshihiro Yamazaki}
\affiliation{Department of Physics, School of Advanced Science and Engineering, Waseda University, Shinjuku, Tokyo 169-8555, Japan}



\begin{abstract}
We investigate the statistical properties of adjacency relationships in a two-dimensional Vicsek model.
We define adjacent edges for all particles at every time step by (a) Delaunay triangulation and (b) Euclidean distance, and obtain cumulative distributions $ P(\tau) $ of 
lifetime $ \tau $ of the edges.
We find that the shape of $ P(\tau) $ changes from an exponential to a power law depending on the interaction radius, which is a parameter of the Vicsek model.
We discuss the emergence of the power-law distribution from the viewpoint of first passage time problem for a fractional Brownian motion.
\end{abstract}


\maketitle

%
\section{Introduction}
Lifetime distributions are commonly utilized in studies aiming to statistically characterize a wide range of stochastic processes in physical and social systems.
There are various examples in which lifetime (or waiting time) distributions exhibit exponential decay: arrival of telephone calls or e-mails, decay of radioactive elements, occurrence of car accidents, and scoring of competitive sports \cite{Merritt2014, Clauset2015}.
Such systems can be theoretically described by a homogeneous Poisson process in which each event occurs independently at a constant rate within a certain time interval \cite{Daley2007}.
Meanwhile, power-law distributions for lifetime are also ubiquitous in nature, because they are associated with the dynamics of earthquakes, solar flares, animal movements, and human activities \cite{Karsai2018}.
Power-law behaviors are often referred to as ``burstiness'' especially for human activities.
A number of stochastic models exhibiting such power-law behaviors have been developed based on an extended Poisson process; examples include the priority queue model \cite{Barabasi2005}, Hawkes process \cite{Hawkes1971}, and cascading Poisson process \cite{Malmgren2008}.

Human activities can be divided into individual-driven and contact-driven (or communication-driven) \cite{Karsai2018}.
Recently, for contact-driven activities, experiments using wearable sensors have been conducted in scientific conferences \cite{Hui2005, Zhao2011}, schools \cite{Stehle2011, Fournet2014, Mastrandrea2015}, companies \cite{Takaguchi2011}, and other settings \cite{Cattuto2010, Isella2011}.
These studies measured how long two people were in close proximity within a certain distance, i.e., the lifetime of the adjacency relationships.
In these cases, it was found that the lifetime distribution obeys a power law. 
%

Power-law properties can be extracted from human activities and from a more general situation.
In this study, we investigate the statistical properties of adjacency relationships in a two-dimensional Vicsek model, which describes the collective motions of self-propelled particles \cite{Vicsek1995, Ginelli2016}.
We focus on the lifetime $ \tau $ during which the adjacency relationships between two particles exist.
%
%
It is found that the cumulative distributions of $\tau$, $ P(\tau) $, becomes an exponential or a power law depending on the interaction radius in the Vicsek model.

\section{Model}
\begin{figure}
	\centering
	\includegraphics[width=8.8cm]{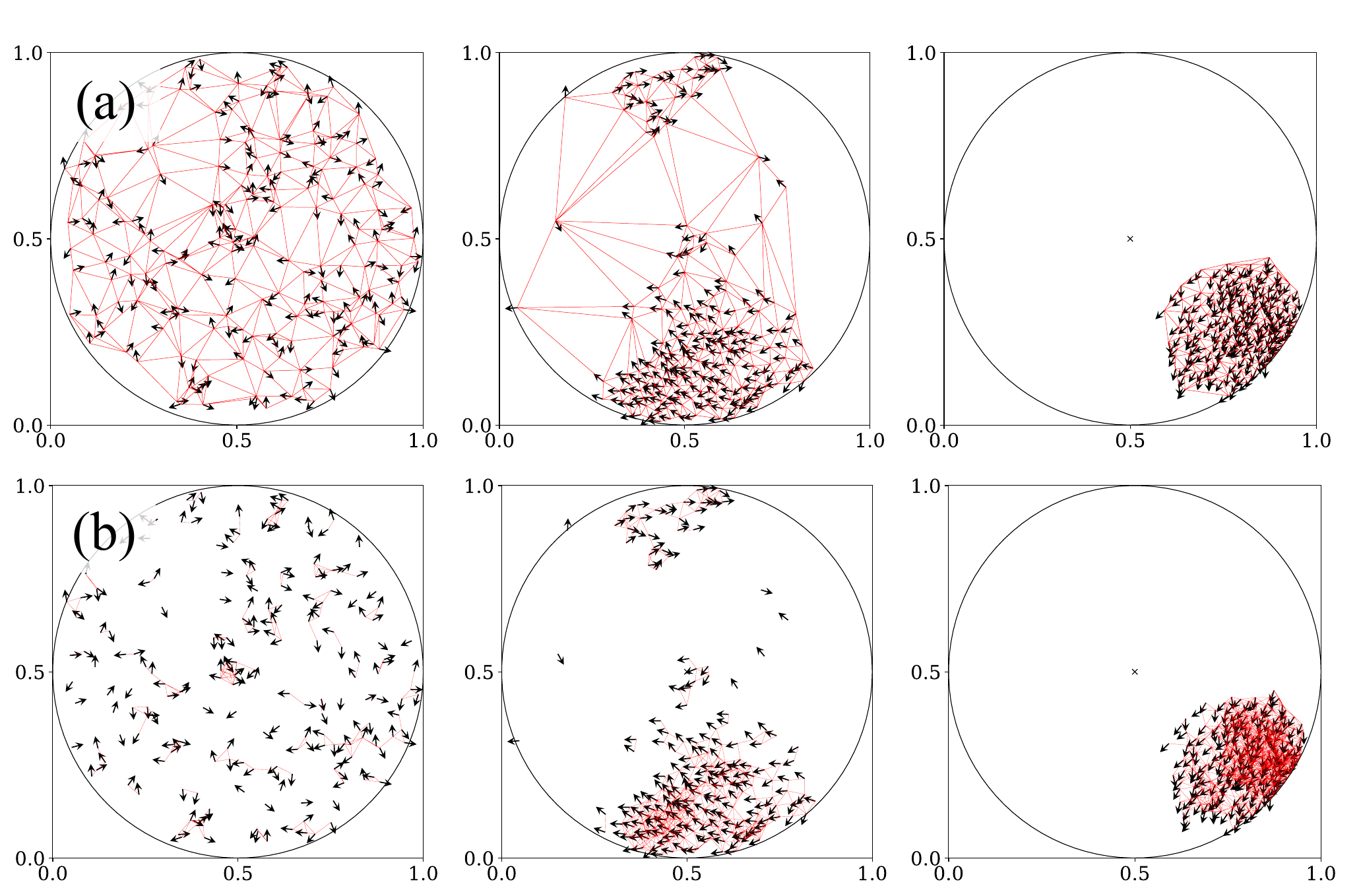}
	\caption{Images of particles' motion and adjacency relationship for $ N=200 $ and $ \eta=0.2 $ at $ t=1100 $ in the circular reflection boundary with diameter $ L=1 $. The direction of the particles is shown by arrows. Lines between particles show the adjacency networks, defined by (a) the Delaunay triangulation and (b) the Euclidean distance $ d < 0.05 $. Left column: $R_{0}=0 $. Particles move randomly because they do not interact. Middle column: $ R_{0}=0.05 $. Particles directions are aligned locally. Right column: $ R_{0}=0.1 $. All particles form a single cluster and move along the boundary.}
	\label{fig:snap_shot}
\end{figure}

Let us consider an $ N $-particle system in a two-dimensional circular space with a diameter $ L $, which corresponds to the system size.
We denote the position and velocity of the $ j $-th particle at time $ t $ as $ \vec{r}_{j}(t) $ and $ \vec{v}_{j}(t) = v_{0} \vec{s}_{j}(t) $, respectively.
Here, $ v_{0} $ is the speed and $ \vec{s}_{j}(t)  $ is the unit vector.
Note that $ \vec{s}_{j}(t)  $ is determined by the angle $ \theta_{j}(t) $ in polar coordinates.
The equation of the motion of each particle is given as follows \cite{Chate2008}:
\begin{align}
	\theta_{j}(t+\Delta t) \nonumber\\ 
	&\hspace{-1.cm}=\mathrm{Arg} \sum_{k\sim j}^{N}\left[(1-c) \vec{v}_{k}(t) + c \vec{f}_{jk}(t) \right] + \xi_{j}(t),\label{vm_1}\\
	\vec{r}_{j}(t+\Delta t) &= \vec{r}_{j}(t) + v_{0} \Delta t \vec{s}_{j}(t+\Delta t).
	\label{vm_2}
\end{align}
In the first term on the right-hand side of Eq. \eqref{vm_1}, the notation $ k\sim j $ in the summation indicates that the $ j $-th particle interacts with others within the circle of radius $ R_{0} $, whose center is $ \vec{r}_{j} $.
In this summation, the first and the second terms show alignment and repulsive interactions between $j$-th and $k$-th particles, respectively.
Here, $ \vec{f}_{jk}(t) $ is given by
\begin{align*}
	\vec{f}_{jk} &= - \vec{e}_{jk} \times \left[1 + \exp \left(\frac{|\vec{r}_{k} - \vec{r}_{j}|}{R_{f}} - 2\right)\right]^{-1},
\end{align*}
where $ \vec{e}_{jk} $ is the unit vector of $\vec{r}_{k} - \vec{r}_{j}$, and $ R_{f} $ is the typical repulsion distance \cite{Chate2008}.
The proportion of the alignment and repulsive interactions is controlled by the parameter $ c $.
The operator $ \mathrm{Arg} $ converts the vector to the angle.
The second term in Eq. \eqref{vm_1} represents noise, or fluctuation, where $ \xi $ is given as a uniform random number in $ [-\eta\pi, +\eta\pi]$ $ (\eta > 0) $.
At every time step, we define the adjacent edges for all particles using: (a) Delaunay triangulation and (b) Euclidean distance $ d $.
Hereafter, these edges are referred to as Delaunay and Euclidean edges, respectively.
The Delaunay triangulation is obtained from the adjacency relationships in the Voronoi regions of each particle (a Voronoi region of a particle is a set of locations, for which the distance to the particle is less than to any other \cite{Okabe1992}).
For the Euclidean edges, we define two particles as adjacent if the Euclidean distance between them is less than $ d $.
Thus, at each time step, particles form an adjacency network as shown in Fig. \ref{fig:snap_shot}.
We focus on the lifetime $ \tau $ during which the adjacency relationship between two particles exist (i.e., the lifetime of adjacent edges).
The adjacency relationships between particles do not affect the particles' motion.
The numerical calculation of Eqs.(\ref{vm_1}) and (\ref{vm_2}) was performed in the circular reflection boundary by setting $ L=1 $, $ \Delta t = 1 $, $\ v_{0}=0.005$, $R_{f}=0.003 $, $ c=0.5 $, and $ d=0.05 $.
$ R_{0} $, $ \eta $, and $ N $ are the controlling parameters.
The total time step was set as $ T=11000 $, and the cumulative distribution of the lifetime $ \tau $, denoted by $ P(\tau) $, is computed using the data of $ \tau $, which are obtained from all adjacent edges for $ t > 1000 $.
\section{Result}
\begin{figure}
	\includegraphics[width=8.8cm]{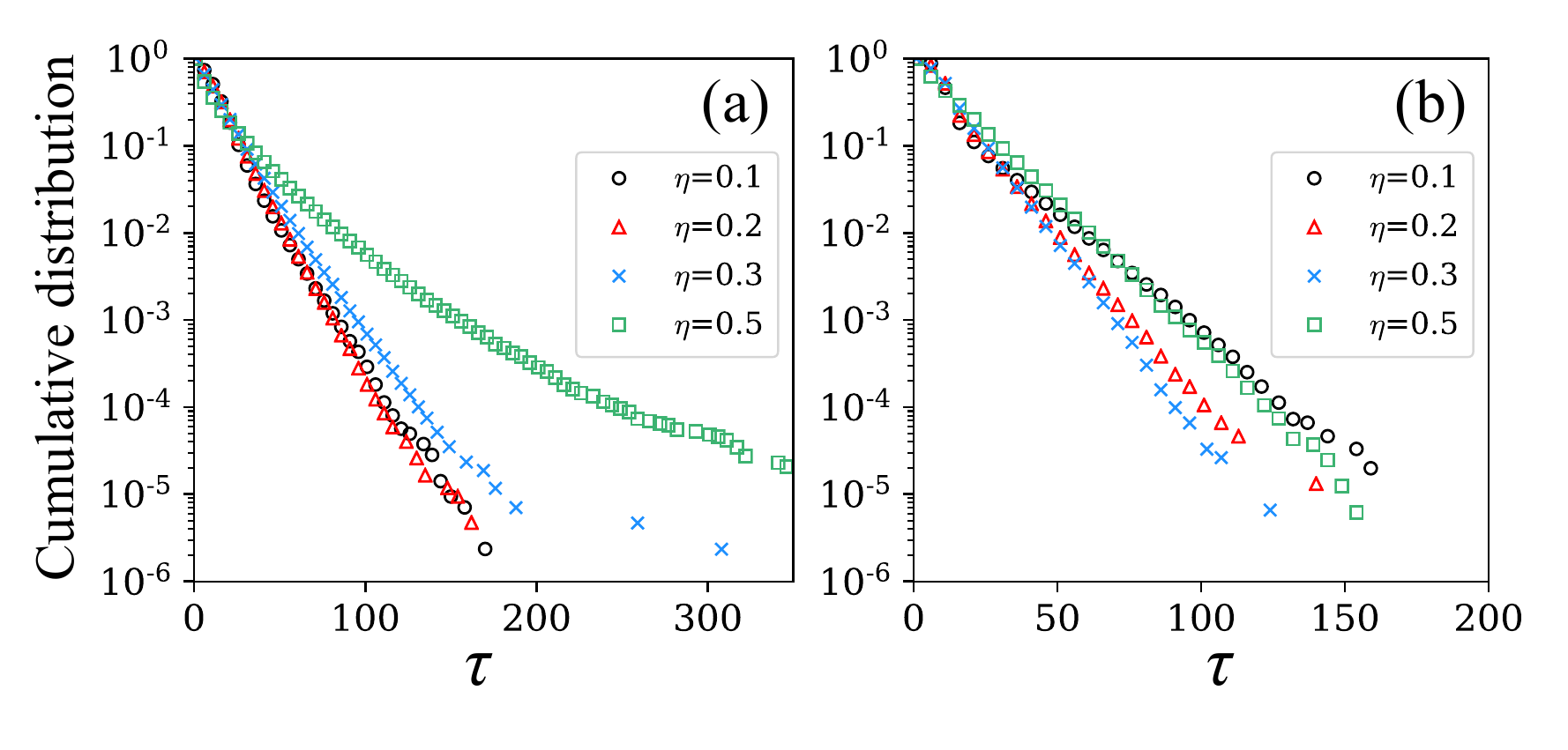}
	\caption{Cumulative distributions $ P(\tau) $ for $ R_{0}=0 $ obtained 
	from (a) Delaunay and (b) Euclidean edges 
	in a semilogarithmic scale. For small $ \eta $, exponential decay is observed.}
	\label{fig:felt0}
\end{figure}

\begin{figure*}
	\includegraphics[width=15cm]{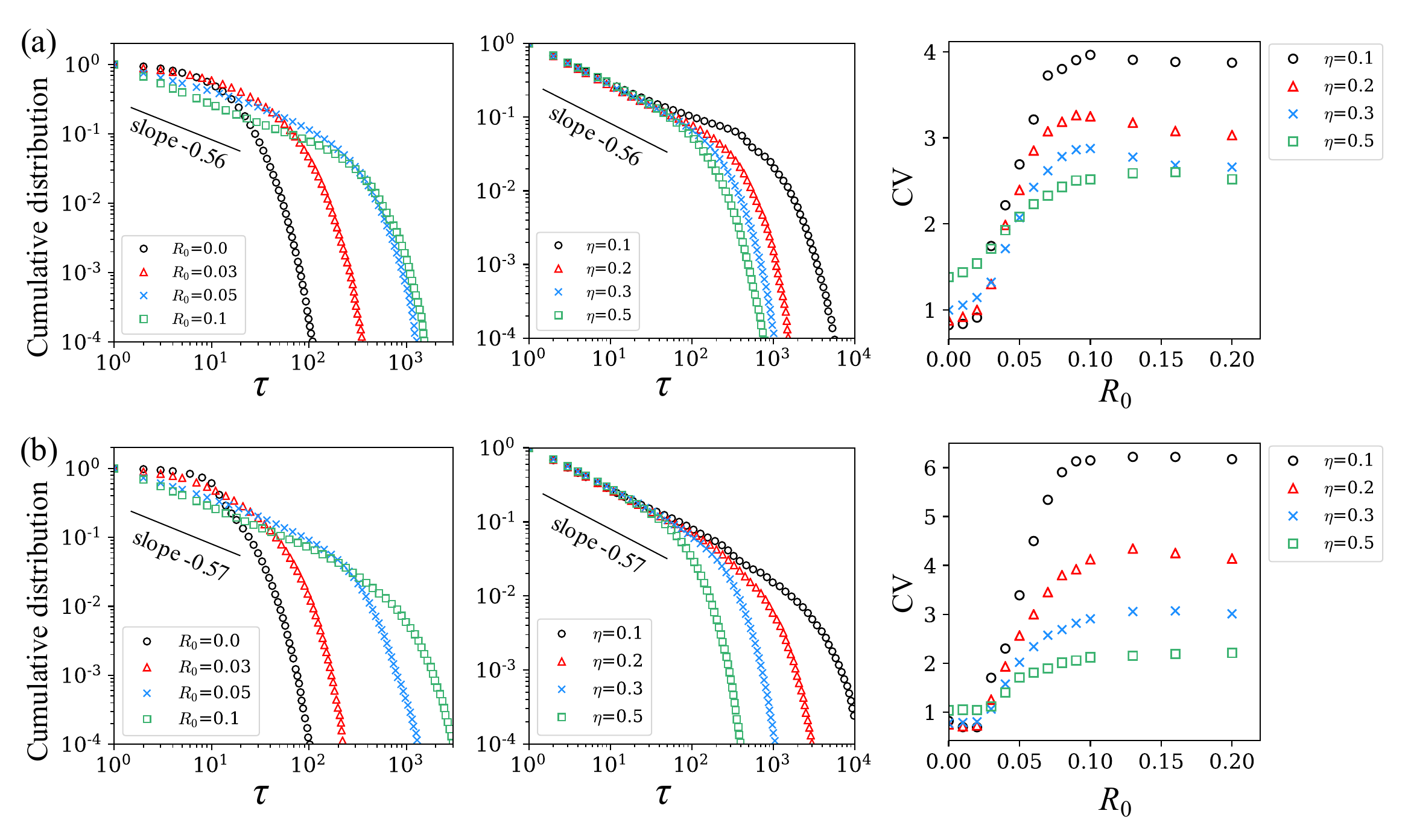}
	\caption{Cumulative distributions $ P(\tau) $ for (a) Delaunay and (b) Euclidean edges. Each panel is shown in a double logarithmic scale. The slope of guideline in each panel is obtained by fittings for $ R_{0}=0.1 $ and $ \eta=0.2 $. Left column: $ R_{0} $  dependence of $ P(\tau) $ where $ \eta=0.2 $; $ P(\tau) $ changes from an exponential to a power-law distributions with an increase in $ R_{0} $. Middle column: $ \eta $ dependence of $ P(\tau) $ where $ R_{0}=0.1 $. Right column: $ R_{0} $ dependence of CV values.}
	\label{fig:felt}
\end{figure*}

We first present results for $ R_{0}=0 $.
In this condition, each particle moves randomly because no interactions occur among them (see the directions of the particles in the left column of Fig. \ref{fig:snap_shot}).
Figure \ref{fig:felt0} shows the cumulative distribution $ P(\tau) $ for (a) Delaunay and (b) Euclidean edges in a semilogarithmic scale.
It is found that $ P(\tau) $ for small $ \eta $ exhibits an exponential decay, although it deviates from the exponential for large $ \eta $.
Next, Fig. \ref{fig:felt} shows $ P(\tau) $ for $ R_{0} \neq 0 $ obtained from (a) Delaunay and (b) Euclidean edges in a double logarithmic scale.
%
%
The left panels in Fig. \ref{fig:felt} show that $ P(\tau) $ follows the power-law distribution 
with an increase in $ R_{0} $.
%
%
The power-law exponents for $R_{0} = 0.1$ and $\eta=0.2$ are estimated as $\alpha \simeq  1.56 $ and 1.57 for the Delaunay and the Euclidean edges, respectively.
The middle panels in Fig. \ref{fig:felt} show the power-law behaviors of $ P(\tau) $ when $ R_{0}=0.1 $ and $ \eta \leq 0.5 $.
We also confirmed that these power-law behaviors are almost independent of $ N ( \gtrsim 10 )$ for both the Delaunay and Euclidean edges.
%

\begin{figure}
	\centering
	\includegraphics[width=8.8cm]{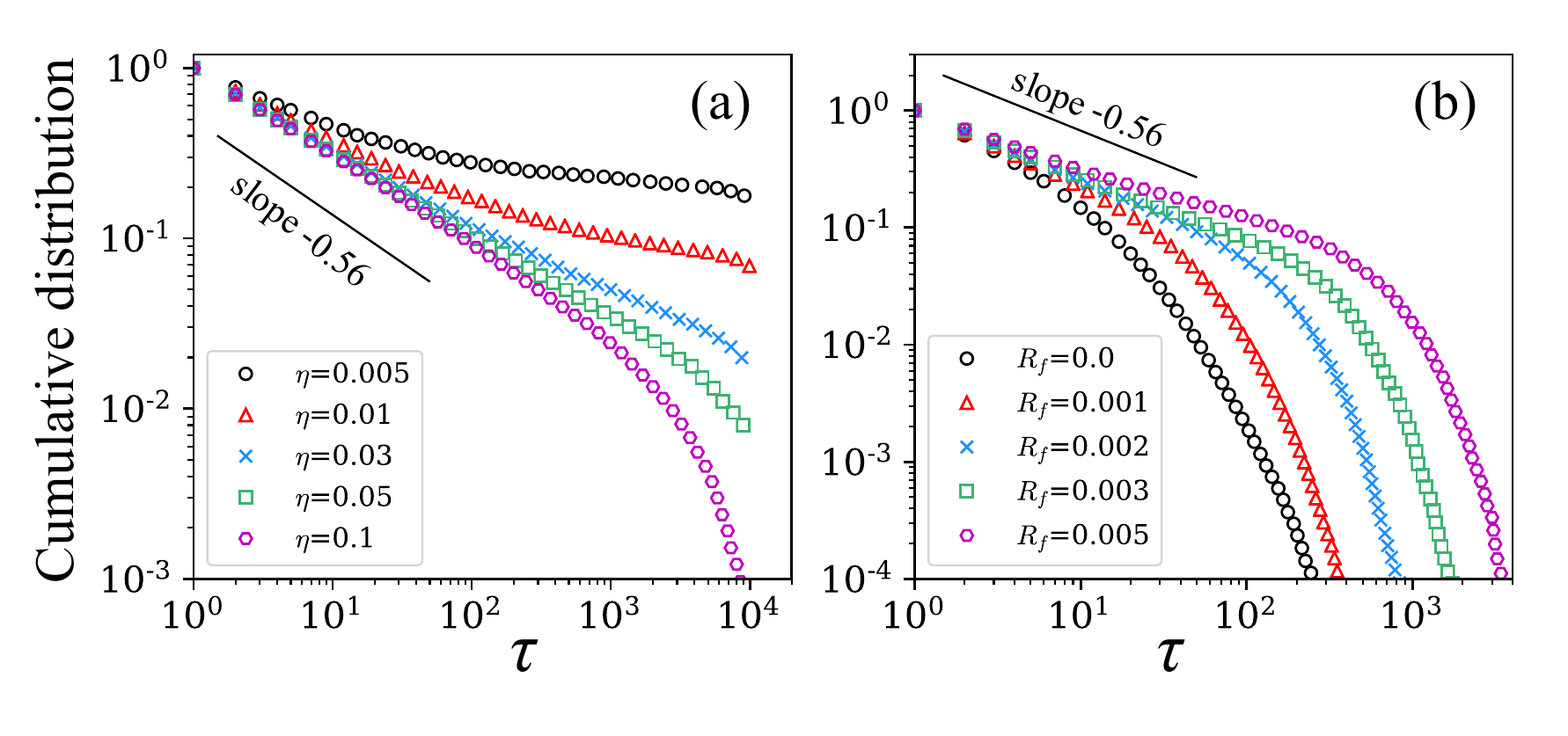}
	\caption{(a) $ \eta $ dependence of $ P(\tau) $ obtained from Delaunay edges 
	without boundaries where $ R_{0}=0.2 $. (b) $ R_{f} $ dependence of $ P(\tau) $ for Delaunay edges where $ R_{0}=0.1 $ and $ \eta=0.2 $.}
	\label{fig:dep}
\end{figure}

To characterize the behaviors of $P(\tau)$, we introduce a coefficient of variation defined as $ \mathrm{CV}= \sigma(\tau)/ \langle \tau \rangle $, where $ \langle \tau \rangle $ and $ \sigma(\tau) $ are the mean and standard deviation of $ \tau $.
$ \mathrm{CV} $ becomes unity when $ \tau $ follows an exponential distribution.
When CV deviates from unity, the distribution is not an exponential.
We present the $ R_{0} $ dependence of $ \mathrm{CV} $ in the right-hand panels of Fig. \ref{fig:felt}.
It is found that $ \mathrm{CV} $ changes from unity to larger values as $ R_{0} $ increases; 
in particular, $ \mathrm{CV} $ increases rapidly at $ R_{0} \simeq 0.05 $.
Therefore, $ P(\tau) $ has a crossover between an exponential and a power-law distribution at around $ R_{0} \simeq 0.05 $.
When $ R_{0} $ is sufficiently large, particles form a single cluster and move together as shown in the right-hand panels of Fig. \ref{fig:snap_shot}.
To check the influence of reflections of particles at the boundary on $ P(\tau) $, we performed a simulation without any boundaries for $ R_{0}=0.2 $.
Note that a single cluster is kept during this simulation.
Figure \ref{fig:dep}(a) shows that the power-law distribution is observed 
with an increase in $ \eta $.
In the small $ \tau $ region, the same power-law exponent $ \alpha \simeq 1.56 $ is obtained independently of $ \eta $.
%
%
The deviation from the power law for small $ \eta $ suggests that the rewiring of edges seldom occurs because of the fixed positions of particles.
This deviation of $ P(\tau) $ from the power law is also observed under the circular boundary.
Thus, reflective interactions between particles and the boundary does not affect the power-law behavior of $ P(\tau) $.
Here, we note that if there is no boundary, $P(\tau)$ is not stationary because a single cluster eventually breaks down due to the noise.
In this sense, existence of boundary is needed to ensure the stationarity of $ P(\tau) $.
\begin{figure}
	\centering
	\includegraphics[width=8cm]{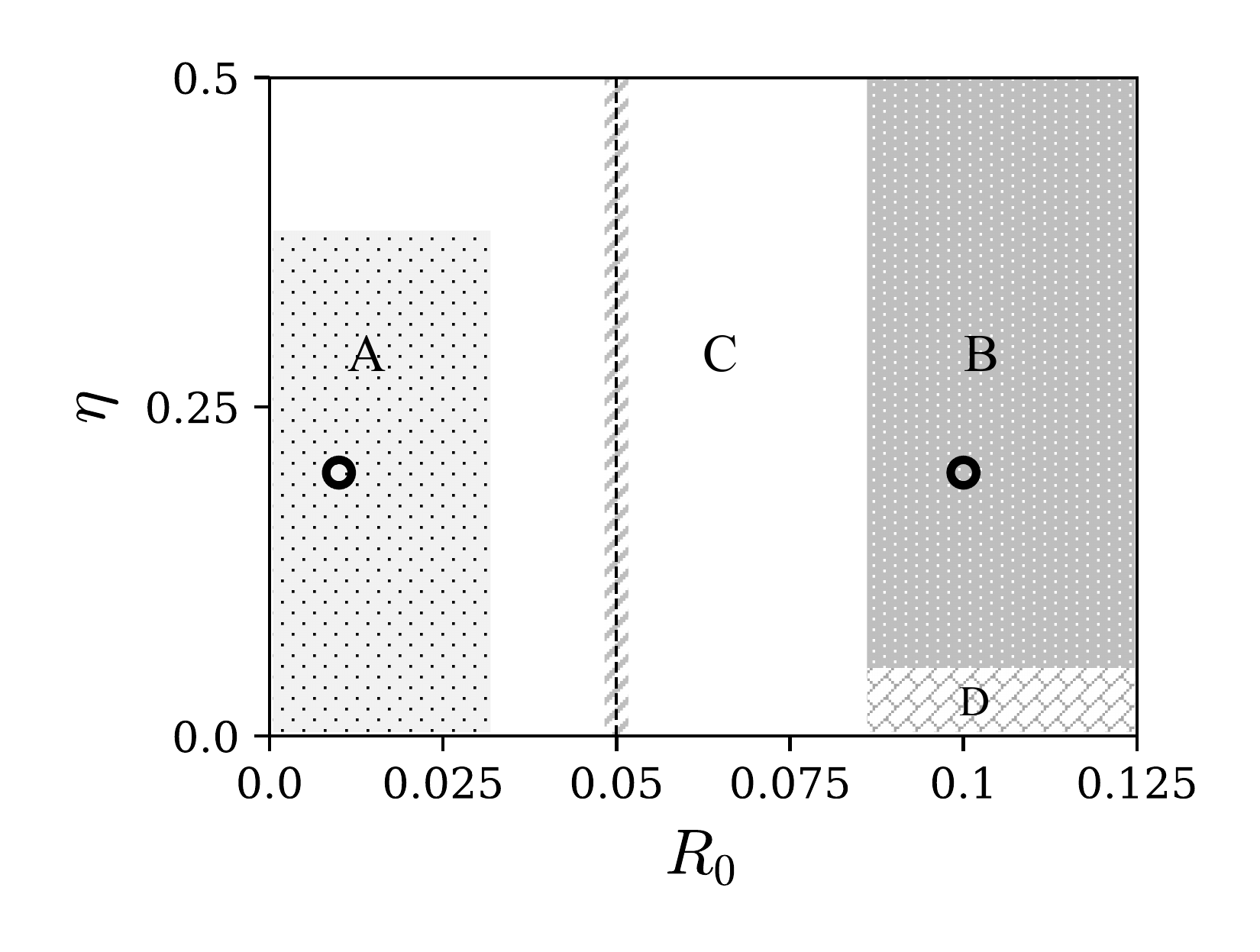}
	\caption{Behaviors of $P(\tau)$ as functions of $ R_{0} $ and $ \eta $. A: exponential; B: power law; C: crossover between A and B; D: $P(\tau)$ becomes a distribution with a longer tail than the power law such as Fig. \ref{fig:dep}(a). The vertical dashed line at $ R_{0} \simeq 0.05 $ represents the point around which $ \mathrm{CV} $ changes considerably.}
	\label{fig:diagram}
\end{figure}
We also check the effect of the repulsive force for large $ R_{0} $ by controlling the repulsion distance $ R_{f} $.
As shown in Fig. \ref{fig:dep}(b), the power-law region with $ \alpha \simeq 1.56 $ reduces with a decrease in $ R_{f} $.
The same result is obtained for the Euclidean edges.
This indicates that the cohesion of particles for a small $ R_{f} $ inhibits emergence of longevity edges.
Thus, the repulsive forces between particles are necessary for the emergence of the power-law behavior of $ P(\tau) $.

As shown in Fig. \ref{fig:diagram}, the typical behaviors of $ P(\tau) $ are classified into four cases A--D.
The exponential decay and power-law of $P(\tau)$ are observed in case A (small $ R_{0} $) 
and case B (large $R_{0}$), respectively. 
The crossover between exponential and power law occurs in case C.
In case D, $P(\tau)$ becomes a distribution with a longer tail 
than the power law, such as Fig. \ref{fig:dep}(a). 
The vertical dashed line at $ R_{0} \simeq 0.05 $ represents the point around which $ \mathrm{CV} $ increases rapidly with an increase in $R_{0}$.


\section{Discussion}
\begin{figure*}
	\centering
	\includegraphics[width=16cm]{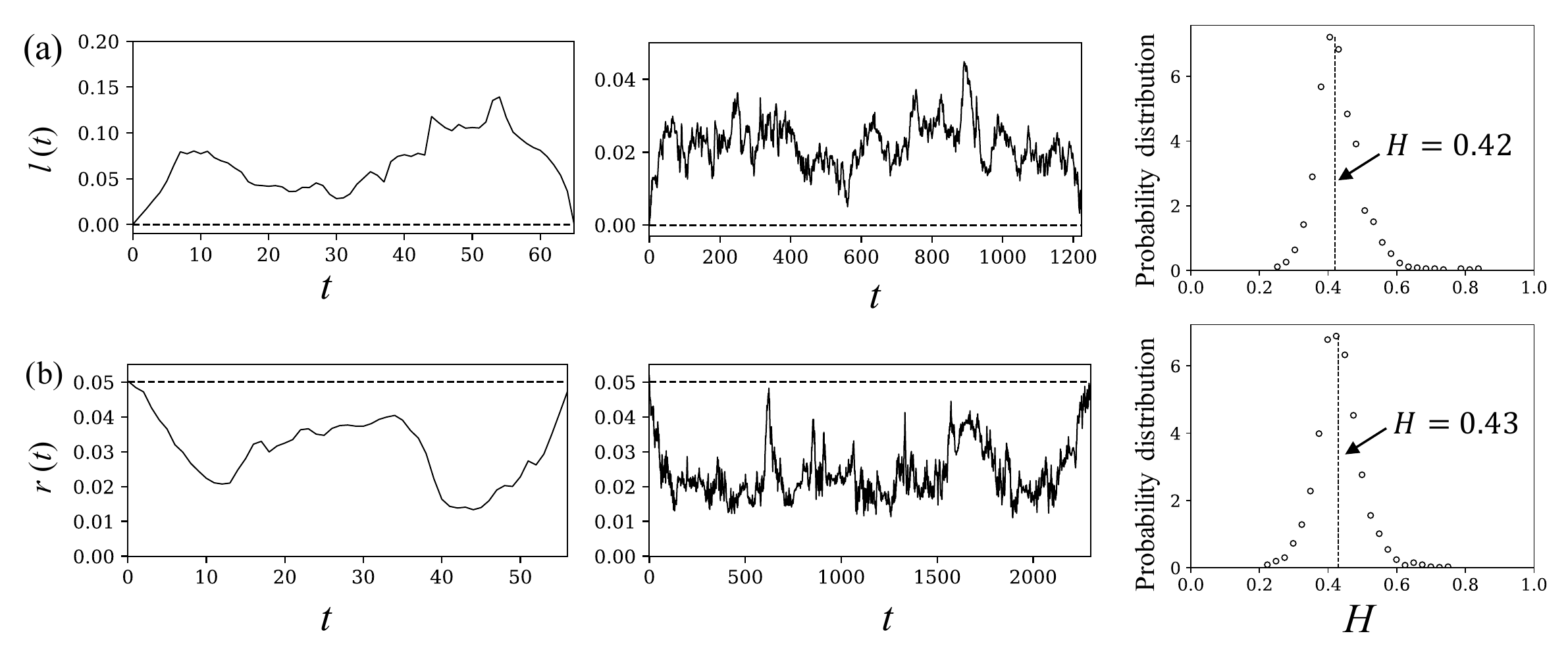}
	\caption{Typical time series of (a) length $ l(t) $ of Voronoi line and (b) distance $ r(t) $ between two particles. Dashed lines indicate $ l(t) = 0 $ and $ r(t)=0.05 $, which show the thresholds for the first return time. Left column: $ R_{0}=0.01 $ and $ \eta=0.2 $ in case A of Fig.\ref{fig:diagram}. Middle column: $ R_{0}=0.1 $ and $ \eta=0.2 $ in case B. Right column: probability distributions of Hurst exponent $ H $ for the time series at $ R_{0}=0.1 $ and $ \eta=0.2 $. Dashed lines indicate the peak values.}
	\label{fig:seq}
\end{figure*}

The lifetime $\tau$ is regarded as a one-dimensional first return time.
For the Delaunay edges, $\tau$ corresponds to the lifetime of the Voronoi line $ l(t) $, where $l(t)$ is the length of each edge.
For the Euclidean edges, $\tau$ is the first return time of the distance $ r(t) $ between two particles to the threshold $ d $, where $ r(t) < d $.
Figure \ref{fig:seq} shows the typical time series of $l(t)$ and $r(t)$, which are obtained at the circle markers in Fig. \ref{fig:diagram}.
It is found that the time series are very different for the two cases.
The power-law behavior of $P(\tau)$ in case B can be explained by considering random fluctuations of $l(t)$ and $r(t)$ as follows.
It is known that the first return time of a fractional Brownian motion follows the power law with the exponent $ \alpha = 2-H $, where $H$ is its Hurst exponent \cite{Ding1995}.
This holds for arbitrary one-dimensional time series characterized by $ H $.
Then, we calculated $ H $ for each time series of $ l(t) $ and $ r(t) $ at $ R_{0}=0.1 $ and $ \eta=0.2 $.
As shown in the right column of Fig. \ref{fig:seq}, $ H $ is distributed  around a peak value, which are $ H\simeq 0.42 $ and $ H\simeq 0.43 $ for the Delaunay and Euclidean edges, respectively.
Because the power-law exponents obtained from $ P(\tau) $ are $ \alpha \simeq 1.56 $ and $ 1.57 $ (see Fig. \ref{fig:felt}), these values satisfy the relation $ \alpha = 2 - H $.
On the other hand in case A, the time series $ l(t) $ and $ r(t) $ have fewer fluctuations and become shorter than those in case B.
The exponential decay of $P(\tau)$ in case A suggests that they are subject to a homogeneous Poisson process.
Previous studies on the Vicsek models have mainly focused on macroscopic properties such as an order--disorder transition of particles' directions or a giant density fluctuation \cite{Ginelli2016}.
In contrast, adjacency relationships of the particles have not been sufficiently examined thus far.
Thus, our present results are novel for the statistical properties in the Vicsek model.
The adjacency relationship of particles is a more detailed characterization of collective motions rather than the macroscopic quantities such as order parameters.
For example, we applied the Delaunay triangulation method to the formation analysis of team sports, i.e., football games \cite{Narizuka2017}.
Nagy {\it et al.} elucidated a hierarchical structure of flocks of birds using the network defined by the time delay between two birds' directions \cite{Nagy2010}.
%
%
We expect that our results can be observed in some experiments, such as those for bacterial motions in circular pools \cite{Wakita2015} and self-propelled robots \cite{Deblais2018}.
\section{Conclusion}
For the lifetime distributions $ P(\tau) $ of the Delaunay and Euclidean edges obtained by the Vicsek model, there exists a crossover for the shape of $ P(\tau) $ between the exponential for small $ R_{0} $ and the power law for large $ R_{0} $.
The power-law exponent $ \alpha $ of $P(\tau)$ satisfies the relation $ \alpha = 2 - H $, where $ H $ is the Hurst exponent obtained from the one-dimensional time series of the Delaunay and Euclidean edges.
\section*{Acknowledgments}
We thank Jun-ichi Wakita, Yuhei Yamada, and Ken Yamamoto for fruitful discussions.
The present work was partially supported by a Grant-in-Aid for Young Scientists No.18K18013 from the Japan Society for the Promotion of Science (JSPS).

\bibliography{reference}

\end{document}